\documentclass{article}

\usepackage{arxiv}

\usepackage[utf8]{inputenc} % allow utf-8 input
\usepackage[T1]{fontenc}    % use 8-bit T1 fonts
\usepackage{hyperref}       % hyperlinks
\usepackage{url}            % simple URL typesetting
\usepackage{booktabs}       % professional-quality tables
\usepackage{amsfonts}       % blackboard math symbols
\usepackage{nicefrac}       % compact symbols for 1/2, etc.
\usepackage{microtype}      % microtypography
\usepackage{lipsum}
\usepackage{graphicx}
\usepackage{xltabular}
\usepackage{threeparttable}

\graphicspath{ {./images/} }

\title{Towards Scalable AASIST: Refining Graph Attention for Speech Deepfake Detection}

\author{
 Ivan Viakhirev \\
  Speechka.ai\\ Information Technologies,\\ Mechanics and Optics University \\
  St. Petersburg,  Russian Federation \\
  \texttt{i.viakhirev@mail.ru} \\
   \And
 Daniil Sirota \\
 St. Petersburg State University\\
 St. Petersburg,  Russian Federation \\
  \texttt{ds11011@yandex.ru} \\
  \AND
 Aleksandr Smirnov \\
Speechka.ai\\ Information Technologies,\\ Mechanics and Optics University \\
  St. Petersburg,  Russian Federation \\
  \texttt{alexsmirnov1213@mail.ru} \\
   \And
   Kirill Borodin\thanks{Corresponding author} \\
  Moscow Technical University \\ of Communications and Informatics\\
  Moscow, Russian Federation\\
  \texttt{k.n.borodin@mtuci.ru}
}

\begin{document}
\maketitle
\begin{abstract}
Advances in voice conversion and text-to-speech synthesis have made automatic speaker verification (ASV) systems more susceptible to spoofing attacks. This work explores modest refinements to the AASIST anti-spoofing architecture. It incorporates a frozen Wav2Vec 2.0 encoder to retain self-supervised speech representations in limited-data settings, substitutes the original graph attention block with a standardized multi-head attention module using heterogeneous query projections, and replaces heuristic frame-segment fusion with a trainable, context-aware integration layer. When evaluated on the ASVspoof 5 corpus, the proposed system reaches a 7.6\% equal error rate (EER), improving on a re-implemented AASIST baseline under the same training conditions. Ablation experiments suggest that each architectural change contributes to the overall performance, indicating that targeted adjustments to established models may help strengthen speech deepfake detection in practical scenarios. The code is publicly available at \url{https://github.com/KORALLLL/AASIST_SCALING}
\end{abstract}

% keywords can be removed

\section{Introduction}
\subsection{Background}
Voice antispoofing research emerged as a direct response to the rapid progress of text-to-speech (TTS) and voice-conversion (VC) systems, which can fabricate speech that fools both humans and automatic speaker-verification (ASV) back-ends\cite{jung2021aasistaudioantispoofingusing, zhang2024improvingshortutteranceantispoofing}. Early countermeasures (CMs) relied on hand-crafted cepstral or constant-Q features and shallow classifiers, but these pipelined designs struggled to generalise beyond the codec and channel conditions seen during training.

The field pivoted toward fully end-to-end architectures once RawNet2\cite{tak2021endtoendantispoofingrawnet2} showed that a single convolutional encoder operating on raw waveforms could learn spoof-specific artefacts without any feature engineering. Shortly afterwards, graph neural networks (GNNs) were introduced to model long-range relations among spectral and temporal cues; RawGAT-ST\cite{tak2021endtoendspectrotemporalgraphattention} demonstrated that spectro-temporal graph attention and competitive pooling could cut equal-error-rate (EER) below 1\% on the ASVspoof 2019 logical-access benchmark\cite{wang2020asvspoof2019largescalepublic}.

Building on these ideas, the AASIST\cite{jung2021aasistaudioantispoofingusing} family unified spectral and temporal graphs inside a heterogeneous stacking graph attention layer (HS-GAL), achieving state-of-the-art performance with a single model. AASIST2\cite{zhang2024improvingshortutteranceantispoofing} inserted multi-scale Res2Net\cite{Gao_2021} blocks and dynamic-chunk training, markedly reducing the short-utterance error rate that had plagued prior systems. Most recently, AASIST3\cite{borodin2024aasist3kanenhancedaasistspeech} augmented every attention block with Kolmogorov-Arnold networks (KANs)\cite{liu2025kankolmogorovarnoldnetworks} and coupled the graph backbone to self-supervised Wav2Vec 2.0 features\cite{baevski2020wav2vec20frameworkselfsupervised}, halving the minimum detection-cost function (minDCF) on the ASVspoof 2024 development set\cite{wang2024asvspoof5crowdsourcedspeech}. While graph attention remains dominant, alternative parameter-efficient architectures are emerging that forego explicit graph reasoning altogether.

Borodin et al.\cite{Borodin2024CapsulebasedAT} introduced ResCapsGuard, which replaces graph attention with dynamic-routing capsules and reaches 2.27\% EER, and Res2TCNGuard, which couples dilated temporal-convolutional branches with a SincNet front-end to achieve 1.49\% EER under 200k parameters. These results confirm that task-specific inductive biases - hierarchical capsules or long-context TCNs - can rival heavier graph or transformer models while remaining edge-deployables.

Recent work converges on self-supervised speech encoders as the de-facto front-end for antispoofing. Fine-tuning only the lower Transformer blocks of WavLM-Base and pooling them with multi-head factorised attention already delivers sub-1\% EER on ASVspoof 2021 DF\cite{Stourbe_2024}. A complementary selective-layer strategy over XLS-R paired with a shallow stacked-linear classifier attains 1.92\% EER and strong out-of-domain generalisation\cite{zhang2024audio}. To further shrink the footprint, state-space sequence models have begun to replace self-attention: a dual-column XLSR-Mamba matches small Transformers on ASVspoof 2019 with roughly one-third fewer parameters\cite{xiao2025xlsrmambadualcolumnbidirectionalstate}. These advances crystallised in ASVspoof 5, whose crowdsourced, codec-rich corpus exposed brittle legacy CMs and showed that SSL front-ends coupled with aggressive codec/noise augmentation can cut minDCF to 0.075 in the open condition\cite{chen2024ustckxdigitdescriptionasvspoof5challenge}.

\subsection{Motivation and problem setting}

Automatic speaker-verification countermeasures have largely stood still since the release of AASIST’s\cite{jung2021aasistaudioantispoofingusing} heterogeneous spectro-temporal graph attention backbone in 2022. Follow-up variants such as AASIST-L\cite{jung2021aasistaudioantispoofingusing}, AASIST2\cite{zhang2024improvingshortutteranceantispoofing} and AASIST3\cite{borodin2024aasist3kanenhancedaasistspeech} prune channels, add multi-scale residual blocks or swap activation functions, yet none revisits the core graph topology, pairwise attention or heuristic max-fusion operator. At the same time, recent ASVspoof challenges\cite{wang2024asvspoof5crowdsourcedspeech} show that compact self-supervised models and codec-aware augmentation can already surpass the unchanged backbone, suggesting that AASIST’s original design may now be over-engineered and brittle in the face of modern deepfakes.

Our study launches a controlled re-examination of AASIST. By confining every experiment to a single benchmark we eliminate cross-dataset variability and concentrate on how targeted architectural tweaks influence detection accuracy. This compact scope allows rapid iteration and provides a clean empirical foundation for subsequent, larger-scale investigations. Our investigation is guided by three research questions:

\textbf{RQ1  -  Frozen SSL Encoders in Data-Constrained Regimes}
How does freezing a self-supervised speech encoder (e.g., Wav2Vec 2.0) influence anti-spoofing performance when the training material is limited to a single corpus such as ASVspoof 5?  

\textit{Rationale.}  Pre-trained SSL representations embed rich acoustic priors that can stabilise optimisation and reduce overfitting, yet a fully frozen front-end may under-specialise to spoof artefacts present in the target domain. Evaluating this trade-off clarifies whether expensive end-to-end fine-tuning is strictly necessary for small-data scenarios.

\textbf{RQ2  -  Standard vs. Pairwise Graph Attention}
Does replacing the bespoke pairwise graph-attention blocks in AASIST with conventional multi-head attention (MHA) lower model complexity and simplify optimisation without degrading detection accuracy?

\textit{Rationale.}  MHA is ubiquitous, well-supported by modern hardware, and easier to maintain. Demonstrating parity or superiority over the specialised pairwise mechanism would justify migrating to a simpler, more portable architecture.

\textbf{RQ3  -  Trainable Fusion vs. Heuristic Max-Pooling}
Can an attention-based fusion layer recover complementary spectral and temporal cues that the original heuristic max-graph operator may discard?  

\textit{Rationale.}  Max-pooling selects only the most salient node activations, potentially ignoring informative but sub-maximal evidence. A learnable fusion scheme might preserve a richer combination of cues, leading to better generalisation against diverse spoofing attacks.

\section{Methodology}\label{methodology}

\subsection{Baseline}
Our baseline couples the Wav2Vec 2.0 XLS-R encoder that transforms raw waveforms into 1024-dimensional contextual embeddings with a MLP  adapter, whose output feeds the original AASIST back-end. AASIST first applies six ResNet blocks, then bifurcates the resulting tensor into parallel spectral and temporal streams that are processed independently before being reunified by a heterogeneous stacking graph-attention layer (HS-GAL), a graph pooling stage, a second HS-GAL, and the max-graph-out (MGO) fusion module; the fused representation is finally passed to a linear classification head that predicts bona-fide versus spoof labels. Throughout the entire stack - from the MLP adapter and ResNet blocks to the HS-GAL and classifier - we replace the original SELU activations with Gaussian Error Linear Units (GELU) to harmonise the non-linearity with the Transformer conventions of the XLS-R front-end; GELU’s smooth, stochastic gating improves gradient flow in attention-heavy networks and yields more stable optimisation without altering the parameter count.

To enhance the model's performance in detecting sophisticated audio manipulations, we introduced a series of architectural and procedural enhancements. These modifications were designed to improve feature representation, optimize information flow, and increase robustness against real-world acoustic conditions.

\subsection{Encoder Freezing (RQ1)}\label{rq1_formulations}

Wav2Vec 2.0 XLS-R remains a fixed feature extractor in all our experiments. Freezing its parameters preserves the rich acoustic priors learned from thousands of hours of unlabelled speech and removes over 300 M parameters from the optimisation graph, leading to faster convergence and greater training stability in our data-constrained setting. Because the ASVspoof 5 training partition is two to three orders of magnitude smaller than the pre-training corpus, allowing gradients to flow through the backbone would risk catastrophic forgetting and inject corpus-specific bias that the downstream anti-spoofing layers cannot easily counteract.

This design choice is specific to the small-data regime studied here. Weight freezing guarantees a stationary, high-quality embedding space for the lightweight adapter and classifier, but it also caps the model’s capacity to specialise. When substantially more labelled or in-domain material is available, progressively unfreezing lower Transformer blocks - or the entire encoder - often yields further gains by letting the network adapt its representations to spoof-specific artefacts. Our results should therefore be interpreted as a lower-bound baseline: they demonstrate that strong performance is achievable without touching the SSL backbone, yet they do not preclude additional improvements from selective fine-tuning in larger-scale scenarios.

\subsection{Revised Graph Attention (RQ2)}\label{rq2_formulation}

The original AASIST backbone employs a bespoke pair-wise graph-attention operator inside each spectral and temporal branch, followed by a heterogeneous stacking graph-attention layer that reunites the two streams. To determine whether such specialised mechanisms are still necessary, we re-implement every attention block as canonical multi-head self-attention (MHA). Within each branch, every node attends to all others through parallel heads, allowing the model to capture diverse relational patterns while exploiting the highly optimised kernels available for Transformer layers. This substitution removes the custom pair-wise computations, lowers implementation complexity, and accelerates both training and inference without altering the input or output dimensionality of the surrounding ResNet blocks.

For cross-modal fusion we preserve the heterogeneous nature of the graph but refine it with a lightweight, type-aware variant of MHA. Separate linear projections generate queries and keys for temporal and spectral nodes, enabling the model to learn modality-specific interaction rules, whereas a single shared projection produces the value embeddings to avoid redundant parameter growth. By standardising the attention formalism yet retaining modality awareness, the revised layer maintains AASIST’s ability to exchange complementary cues across time and frequency while benefiting from the stability and efficiency of uniform self-attention.

\subsection{Trainable Fusion (RQ3)}\label{rq3_formulation}
The original AASIST backbone merges its spectral and temporal branches with a heuristic torch.max operation that selects the element-wise dominant activation. While computationally cheap, this hard selection imposes a narrow information bottleneck: any feature that is salient in one stream but sub-maximal in the other is irrevocably discarded, and gradient flow is restricted to whichever branch "wins" at each position. Such behaviour is ill-suited to anti-spoofing, where complementary cues often manifest with different magnitudes across time- and frequency-centric representations.

We therefore replace the max operator with a fully learnable multi-head self-attention (MHA) fusion module. The spectral and temporal embeddings are first concatenated along the node dimension, after which parallel attention heads generate query-key-value projections that let every node attend to the joint context of both modalities. Because attention weights are learned end-to-end, the model can amplify subtle but informative patterns that might be overshadowed in a max scheme, while simultaneously down-weighting redundant or noisy activations. The resulting fused tensor preserves the original dimensionality, enabling a drop-in substitution that leaves the downstream classifier unchanged.

\section{Experimental setup}\label{experimental_setup}

All experiments are carried out on the ASVspoof 5 corpus\cite{wang2024asvspoof5crowdsourcedspeech}. Audio is resampled to 16 kHz and fed to the model as single-channel, 32-bit floating-point tensors. Training runs on four NVIDIA V100 GPUs with synchronous data parallelism; the global mini-batch contains forty-eight utterances (twelve per GPU). Optimisation lasts twenty epochs and employs the NAdam optimiser\cite{dozat.2016} with a learning rate of $1\times10^{-4}$. A cosine-annealing scheduler with $T_{max}=300$ steps modulates the learning rate, and the first two epochs omit validation to accelerate the warm-up phase.

The data-augmentation pipeline is implemented in \texttt{torch-audiomentations}\cite{jordal_torch_audiomentations_2025} and rebuilt on-the-fly whenever its global application probability $p$ or intensity coefficient $\kappa$ is updated. Both variables are linearly annealed during the first ten epochs, starting at $p_0=0.5$ and $\kappa_0=1.0$ and reaching $p_{max}=0.9$ and $\kappa_{max}=1.8$. Each call first selects exactly one “codec” corruption with probability 0.4: MP3 compression with a bitrate that shrinks from 128-64 kb/s down to 64-32 kb/s as $\kappa$ grows, telephone-bandwidth filtering, resampling-based low-bitrate compression, or additive coloured noise whose signal-to-noise ratio ranges from 20 dB down to 3 dB. The surviving signal then traverses a chain of gain modulation ($\pm6$ dB), pitch shifting up to $\pm 2$ semitones scaled by $\kappa$, random temporal shifts up to ten per-cent of the waveform, and a low-pass filter whose cutoff moves between 2 kHz and 7.5 kHz. Peak normalisation is applied last to standardise amplitude.

RawBoost perturbations\cite{tak2022rawboostrawdataboosting} are sampled independently of the main pipeline. Three primitive algorithms - convolutive noise, impulsive noise, and coloured additive noise - can act individually, serially, or in parallel, yielding eight canonical variants. Each algorithm draws its own parameter set (band-centre frequencies, bandwidths, filter coefficients, noise gains and bias terms) from the ranges proposed by the original RawBoost publication; the selected variant processes every training utterance exactly once per epoch.

We use the data-augmentation pipeline and different RawBoost pertuberations only for training our final setup.

Training uses a Hybrid Loss that interpolates between cross-entropy and focal loss\cite{lin2018focallossdenseobject}. Interpolation begins when the best validation EER first falls below 8\% and progresses linearly over five epochs until focal loss dominates. Focal loss employs $\gamma=2$ and an $\alpha=0.25$ class-weighting scheme to emphasise hard negatives once the classifier has mastered the easy cases. All random number generators are seeded at initialisation, and only the adapter, ResNet blocks, attention layers and classifier receive gradient updates - the 300-million-parameter Wav2Vec 2.0 XLS-R backbone\cite{babu2021xlsrselfsupervisedcrosslingualspeech} remains frozen throughout. Full training cycle completes in approximately sixteen hours on the V100 cluster and requires no more than 16 GB of memory per GPU.

\section{Results and Discussion}\label{sec:results}
\begin{table*}[hbt!]
  \centering
  \begin{threeparttable}
    \caption{Effect of freezing the Wav2Vec\,2.0 front‐end (RQ1)}
    \label{tab:rq1_ablation}
    \begin{tabularx}{\textwidth}{@{}>{\raggedright\arraybackslash}X c@{}}
      \toprule
      \textbf{Configuration} & \textbf{EER (\%)} \\
      \midrule
      Baseline AASIST                            & 27.58 \\
      \midrule
      Trainable Wav2Vec front-end                & 21.67 \\
      Frozen Wav2Vec front-end                   &  8.76 \\
      \midrule
      \textbf{Full Proposed Modifications}       & \textbf{7.66} \\
      \bottomrule
    \end{tabularx}
    \begin{tablenotes}
      \footnotesize
      \item \textbf{Baseline AASIST} is the original six-block ResNet backbone with separate spectral and temporal streams, bespoke graph attention, heuristic max fusion, and no self-supervised pre-training.
      \item \textbf{Trainable/Frozen} rows replace the convolutional front-end with Wav2Vec 2.0 XLS-R 300 M, either fine-tuned end-to-end (trainable) or kept fixed (frozen) during optimisation, as described in Sec. \ref{rq1_formulations}
      \item \textbf{Full Proposed Modifications} add standard multi-head graph attention, learnable attention-based fusion, and the dynamic augmentation schedule from Sec. \ref{experimental_setup}, achieving the lowest equal-error rate.
    \end{tablenotes}
  \end{threeparttable}
\end{table*}

This section analyses the empirical impact of each architectural intervention on the ASVspoof~5 corpora.  All error rates are reported as equal-error rate (EER) percentages; lower values indicate superior spoof-detection performance.

\subsection{RQ1}

Table~\ref{tab:rq1_ablation} confirms that \emph{freezing} the Wav2Vec~2.0 XLS-R~300M backbone is crucial in the data-constrained regime studied.  Replacing AASIST’s sinc-convolutional front-end with a \emph{trainable} SSL encoder lowers the EER from $27.58\%$ to $21.67\%$, but leaves a large performance gap relative to the frozen variant.  Locking the encoder parameters further reduces the EER to $8.76\%$, a three-fold improvement over the fine-tuned counterpart.  These findings support our hypothesis that preserving the broad acoustic priors learned during large-scale pre-training yields more discriminative, spoof-aware representations than attempting end-to-end optimisation on the limited ASVspoof~5 corpus.  They also validate the design choice to freeze the front-end in all subsequent experiments.

\subsection{RQ2}

\begin{table*}[hbt!]
  \centering
  \begin{threeparttable}
    \caption{Effect of replacing bespoke graph attention with multi-head attention (RQ2)}
    \label{tab:rq2_ablation}
    \begin{tabularx}{\textwidth}{@{}>{\raggedright\arraybackslash}X c@{}}
      \toprule
      \textbf{Configuration} & \textbf{EER (\%)} \\
      \midrule
      Baseline AASIST\,{+}\,Frozen Wav2Vec\,(\emph{bespoke graph attention}) & 8.76 \\
      Multi-head self-attention in place of bespoke graph attention          & 8.43 \\
      \midrule
      \textbf{Full Proposed Modifications}                                   & \textbf{7.66} \\
      \bottomrule
    \end{tabularx}
    \begin{tablenotes}
      \footnotesize
      \item \textbf{Baseline.} Original AASIST back-end with six ResNet blocks, separate spectral and temporal streams, bespoke pair-wise graph attention layers, heuristic max fusion, and a \emph{frozen} Wav2Vec 2.0 XLS-R 300 M front-end.
      \item \textbf{Multi-head self-attention.} Replaces every bespoke graph-attention block with standard multi-head self-attention while leaving all other components unchanged, isolating the impact of attention formalism, as described in Sec.\ref{rq2_formulation}.
      \item \textbf{Full Proposed Modifications.} Combine multi-head self-attention with learnable attention-based fusion and the dynamic augmentation schedule from Sec.\ref{experimental_setup}, producing the lowest equal-error rate.
    \end{tablenotes}
  \end{threeparttable}
\end{table*}

The ablation in Table~\ref{tab:rq2_ablation} isolates the effect of replacing AASIST’s bespoke pair-wise graph attention with conventional multi-head self-attention (MHA).  Under an otherwise identical configuration - including the frozen SSL front-end - the swap cuts the EER from $8.76\%$ to $8.43\%$.  Although the absolute gain is modest, the result demonstrates that the lighter, hardware-friendly MHA matches and slightly surpasses the hand-crafted operator without incurring any penalty in detection accuracy.  Coupled with the substantial reduction in implementation complexity, this outcome argues strongly for adopting MHA as the default attention mechanism in future anti-spoofing models in this study.

\subsection{RQ3}
\begin{table*}[hbt!]
  \centering
  \begin{threeparttable}
    \caption{Effect of learnable fusion versus heuristic max pooling (RQ3)}
    \label{tab:rq3_ablation}
    \begin{tabularx}{\textwidth}{@{}>{\raggedright\arraybackslash}X c@{}}
      \toprule
      \textbf{Configuration} & \textbf{EER (\%)} \\
      \midrule
      Baseline AASIST\,{+}\,Frozen Wav2Vec\,{+}\,MHA\,(heuristic \texttt{torch.max} fusion) & 8.43 \\
      Learnable soft fusion implemented with MHA                                          & 7.93 \\
      \midrule
      \textbf{Full Proposed Modifications}                                               & \textbf{7.66} \\
      \bottomrule
    \end{tabularx}
    \begin{tablenotes}
      \footnotesize
      \item \textbf{Baseline.} Original AASIST back-end enhanced with a frozen Wav2Vec 2.0 XLS-R 300 M front-end and standard multi-head self-attention (MHA) inside each graph layer, but still relying on the heuristic element-wise maximum to merge spectral and temporal streams.
      \item \textbf{Learnable soft fusion.} Replaces the heuristic \texttt{torch.max} operator with a trainable MHA-based fusion module that jointly attends over concatenated spectral and temporal embeddings, as described in Sec. \ref{rq3_formulation}.
      \item \textbf{Full Proposed Modifications.} Combine soft fusion with the other architectural refinements and the dynamic augmentation schedule from Sec.\ref{experimental_setup}, achieving the lowest equal-error rate.
    \end{tablenotes}
  \end{threeparttable}
\end{table*}

Table~\ref{tab:rq3_ablation} evaluates the transition from a heuristic \texttt{torch.max} merger to a trainable, MHA-based fusion layer.  Starting from the RQ2 baseline, the learnable module cuts the EER from $8.43\%$ to $7.93\%$, indicating that adaptive weighting of spectral and temporal cues captures complementary evidence that the hard maximum discards.  The improvement is consistent with our intuition that spoof artefacts may manifest with different magnitudes across modalities and therefore benefit from soft, context-aware aggregation.

\subsection{Synergy of All Modifications}

When all three modifications  -  frozen Wav2Vec~2.0, standard MHA in graph blocks, and learnable fusion - are applied simultaneously, the EER drops to $7.66\%$ (\textbf{Full Proposed Modifications} rows in Tables~\ref{tab:rq1_ablation}-\ref{tab:rq3_ablation}).  The composite reduction of $19.92$ absolute points relative to the original AASIST baseline underscores the complementary nature of the interventions.  In particular, most of the gain arises from the front-end freezing (RQ1), while attention formalism (RQ2) and fusion strategy (RQ3) provide additive but non-trivial refinements.  Together, these results validate the research questions posed in Section~\ref{methodology} and demonstrate that careful modernisation of a legacy backbone can yield state-of-the-art performance without increasing model size.

\section{Conclusion}

This study offers an initial re-examination of AASIST on the ASVspoof 5 benchmark. Three lightweight changes - (i) freezing a Wav2Vec 2.0 front-end, (ii) replacing bespoke graph attention with standard multi-head self-attention, and (iii) adding a learnable fusion layer - jointly reduce equal-error rate from 27.58\% to 7.66\%.

The scope is intentionally narrow: one corpus, a single spoofing condition, and a fully frozen encoder. Future work will check whether these gains extend to other anti-spoofing benchmarks, experiment with progressive unfreezing, and explore alternative back-bones. We regard these findings as a first step toward a broader reformulation of AASIST.

\section{Acknowledgements}
This study emerged from the "Summer with AIRI" School of Artificial Intelligence. We are grateful to Alexandra Broytman, Ekaterina Mamontova, and the entire AIRI team for conceiving and running the school, creating the environment that enabled this project, providing V100 resources, and offering steady mentorship throughout.

Our appreciation extends to Mkrtchian Grach for his engaging lecture on voice-antispoofing, which inspired several of the experiments reported here. We also thank Oleg Rogov and Dmitrii Korzh for their illuminating lectures on AI safety and the security of voice biometric systems, which provided valuable context for understanding the broader implications of antispoofing research.

\clearpage
\bibliographystyle{plain}
% \bibliography{references}

\end{document}